\documentclass[11pt]{article}
\usepackage[a4paper,hmarginratio=1:1,vmarginratio=2:3,
totalwidth=15.8cm,totalheight=22.5cm]{geometry}
\usepackage{bm,epstopdf,epsfig,amsmath,amssymb,amsfonts,colordvi,latexsym,comment,cancel,
verbatim}
\usepackage{graphicx,graphics}
\usepackage[font=md,captionskip=8pt]{subfig}
\usepackage[usenames,dvipsnames]{color}

\usepackage{setspace}
\newcommand{\cgs}{c_{\textnormal{\tiny\textsc{GS}}}}

\newcommand{\cA}{{\cal A}}
\newcommand{\cB}{{\cal B}}

\newcommand{\cE}{{\cal E}}
\newcommand{\cF}{{\cal F}}

\newcommand{\cK}{{\cal K}}
\newcommand{\cL}{{\cal L}}

\newcommand{\cO}{{\cal O}}

\newcommand{\cR}{{\cal R}}
\newcommand{\cS}{{\cal S}}

\newcommand{\cV}{{\cal V}}

\newcommand{\cW}{{\cal W}}

\newcommand{\bE}{{\bf E}}

\newcommand{\Tr}{\mbox{Tr}}

\renewcommand{\Re}{\text{Re}\ }
\renewcommand{\Im}{\text{Im}\ }

\newcommand{\ra}{\rightarrow}
\newcommand{\be}{\begin{equation}}
\newcommand{\ee}{\end{equation}}
\newcommand{\bea}{\begin{eqnarray}}
\newcommand{\eea}{\end{eqnarray}}
\newcommand{\Ra}{\Rightarrow}

\newcommand{\otheta}{\overline\theta}
\newcommand{\opsi}{\overline\psi}
\newcommand{\baa}{\begin{array}}
\newcommand{\eaa}{\end{array}}
\long\def\symbolfootnote[#1]#2{\begingroup
\def\thefootnote{\fnsymbol{footnote}}\footnote[#1]{#2}\endgroup}
\begin{document}
\begin{flushright}
CERN-PH-TH-2014-242\\
\end{flushright}

\thispagestyle{empty}

\vspace{2cm}

\begin{center}
{\Large {\bf 
Gauged R-symmetry and its anomalies 

\bigskip
in 4D N=1 supergravity and phenomenological implications
}}
\\
\medskip
\vspace{1.cm}
\textbf
{
I. Antoniadis$^{\,a, b, c}$, 
D. M. Ghilencea$^{\,d, e\,}$,
R. Knoops$^{d,\,f\,}$}

\bigskip

$^a$ {\small Albert Einstein Center for Fundamental Physics, 
 Institute for Theoretical Physics,}

{\small  University of Bern,  5 Sidlestrasse, CH-3012 Bern, Switzerland}

$^b$ {\small  LPTHE, Universite Pierre et Marie Curie, F-75252 Paris, France}

$^c$ {\small Ecole Polytechnique, F-91128 Palaiseau, France}

$^d$ {\small CERN Theory Division, CH-1211 Geneva 23, Switzerland}

$^e$ {\small Theoretical Physics Department, National Institute of Physics and}

{\small  Nuclear Engineering (IFIN-HH) Bucharest, MG-6 077125, 
Romania.}

$^f$ {\small Instituut voor Theoretische Fysica, KU Leuven,
Clestijnenlaan 200D, B-3001 Leuven, Belgium}

\end{center}
\bigskip

\begin{abstract}
\noindent
We consider  a class of models 
with gauged $U(1)_R$ symmetry in 4D  N=1  supergravity 
 that  have, at the classical level, a metastable ground state,
an infinitesimally small (tunable) positive cosmological constant 
and a TeV gravitino mass.  
We analyse if these properties are maintained
under the addition of visible sector (MSSM-like) and hidden sector state(s),
where the latter may be needed for quantum consistency.
We then discuss the anomaly cancellation conditions in supergravity
 as derived by Freedman, Elvang and K\"ors and apply their results to the
 special case of a  $U(1)_R$ symmetry,  
in the presence of the Fayet-Iliopoulos term 
($\xi$) and Green-Schwarz mechanism(s). We investigate the 
relation of these anomaly cancellation conditions to the  ``naive''  field theory 
approach in global SUSY, in which case $U(1)_R$ cannot 
even be gauged.  We show the two approaches give similar conditions.
Their induced constraints at the phenomenological level, on the above models,
 remain strong even if one lifted the GUT-like conditions for the MSSM
gauge couplings.  
In an anomaly-free model, a tunable,  TeV-scale gravitino mass 
may remain possible  provided that  the  $U(1)_R$ charges of additional 
hidden sector fermions  (constrained by the cubic anomaly alone)
do  not  conflict with  the related values of $U(1)_R$ charges of their scalar
superpartners, constrained by  existence of a stable ground state.
This issue may be bypassed  by tuning instead the coefficients of 
the  Kahler connection anomalies ($b_K, b_{CK}$).
\end{abstract}

\newpage

\section{Introduction}

The aim to obtain de Sitter vacua in 4D N=1 supergravity with an infinitesimally
small, positive (tunable) cosmological constant is a  difficult 
task \cite{Maloney:2002rr,Burgess:2003ic,Kachru:2003aw,Choi:2004sx,Choi:2005uz,
Choi:2005ge,Endo:2005uy,Lebedev:2006qq,Banks:2014pqa}.
One possible attempt in this direction is to  use models with a gauged R-symmetry ($U(1)_R$)
\cite{Cremmer:1983yr,Chamseddine:1995gb,Castano:1995ci,Binetruy:2004hh,Lee:2008zs}.
The models with anomalous $U(1)$ gauge symmetries other than $U(1)_R$ 
 usually  lead to anti-de Sitter (AdS) minimum with broken supersymmetry  (SUSY)
\cite{Antoniadis:2008uk} or the positivity of their squared soft scalars masses is 
difficult to achieve \cite{Kawamura:1996wn,Kawamura:1996bd}.

In   \cite{V1,IA} a  minimal toy model based on
a $U(1)_R$ gauge symmetry was studied.
In its original version,  the minimal field content of the model is the supergravity multiplet 
$(e_\mu^i, \psi_{3/2})$ coupled to the gauge multiplet of $U(1)_R$ and the (string) 
dilaton superfield $S$. 
The dilaton $S$ (of scalar component $s$) enables the presence of  a  
shift symmetry, $S\ra S-i \,\cgs\,\Lambda$,  with $\cgs$  a real constant,
as a consequence of the dual representation description in terms of two-index
antisymmetric tensor for $\Im[s]$. 
This symmetry is gauged and $S$ is the only field that transforms non-linearly under it. 
Thus,  the dilaton participates to  the 4D Green-Schwarz (GS) mechanism. 
The only superpotential allowed by this symmetry is of the 
form $W(s)=a \exp(b\,s)$ with  $a$  a 
real constant and $b<0$, with\footnote{The case  $b<0$ is nonperturbative. 
The case $b>0$ is not considered in this work.}
 a Kahler potential  $\cK(s,\overline s)= -2\kappa^{-2}  \ln (s+\overline s)$.  

With this minimal action and field content, 
one can show that there exists a 
 ground state for a scalar field vev with 
$\alpha\equiv b \langle s\!+\!\overline s\rangle\!=\! -0.1833$ \cite{V1,IA}. This ground state 
thus depends on $b$ only, for a vanishing vacuum energy. There is an extremely
mild dependence on $\cgs$ as well,  if one also demands 
an infinitesimally  small positive (tunable) value of the cosmological constant. 
At the same time, the gravitino mass $m_{3/2}\!\sim\! \kappa^{-1} (a\,b)$
is near the TeV scale\footnote{We 
use $\kappa = 1/m_p$,  $m_p = {M_{\text{Planck} } }/(\sqrt{8
 \pi}) = 2.4 \times 10^{18}$ GeV.} (by tuning $a$)  and the scale of supersymmetry 
 breaking is that of gravity mediation $\sqrt{\langle  F^s\rangle}\!=\!
(m_{3/2}\,m_P/\vert b\vert)^{1/2}$
GeV  (by tuning $b$), with both
$F$ and $D$ term breaking from the dilaton.

The  purpose of this paper is to study further this model
with $U(1)_R$ gauge symmetry,  at both classical and quantum levels.
We investigate if these  nice properties of the model can be maintained in 
the  presence of additional states in the visible and hidden sectors that a 
more realistic model demands. This  model is extended  in the  visible sector by  
the gauge and chiral multiplets of the minimal supersymmetric standard model (MSSM).
The effect  of adding states in the  hidden sector that may be 
demanded for  quantum consistency (like anomaly cancellation) is also investigated
while trying to maintain a TeV gravitino mass. Particular attention
is paid to the anomaly cancellation mechanism in supergravity with 
gauged $U(1)_R$.

The plan of the paper is as follows.
In Section~\ref{sec1} we review the initial toy model 
and  examine the metastability of the ground state.  
We then consider two cases: 1) the addition  
 $U(1)_R$-neutral superfield  in the visible sector and
2) the case of a  $U(1)_R$-charged superfield in the hidden sector. We check
 under what conditions the  nice properties of the initial 
ground state that we mentioned (like $m_{3/2}\sim$TeV, etc)  are maintained.

In Section~\ref{anomalies} we
investigate the anomaly cancellation in supergravity with gauged $U(1)_R$.
A general study of anomaly cancellation conditions
 in supergravity with a gauge group $G \times\! U(1)$ and   a Green-Schwarz  mechanism(s)
and Fayet Iliopoulos terms (FI) of the anomalous $U(1)$  was presented in two 
interesting papers  \cite{F1,F2} (see also \cite{dudas1}).
We study the exact relation of these anomaly cancellation conditions 
to the ``naive'' field theory approach conditions in global SUSY 
(using $\Tr$  over charges and GS mechanism), 
for the  special case  of gauged $U(1)_R$. Note that in global SUSY $U(1)_R$ 
cannot  be gauged \cite{west}. The spectrum is that of minimal 4D N=1 supergravity
 extended by MSSM-like superfields with gauge group of Standard Model (SM)
 times $U(1)_R$.
This is an interesting check and a stand-alone result, 
independent of the rest of the paper.

In Section~\ref{sec4} we apply these results to MSSM-like models.  We 
assume  that matter superfields have  $U(1)_R$-charges equal to $Q$,
use the GS mechanism in the presence of  $U(1)_R$ FI terms (which shift
the fermions charges) and, notably,  relax GUT-like unification condition.
Even after doing so, the anomaly cancellation remains a strong parametric constraint.
We then examine, in simple examples, if the gravitino mass remains 
tunable to TeV values.
We  show that the addition to the MSSM hidden sector of a $U(1)_R$  charged superfield
(of fermion component $\psi_z$ and charge $R_z$) does not solve
this problem.
This is because the $U(1)_R$ cubic anomaly-induced  constraint on 
the R-charge $q_z\sim R_z-\xi$ of  the scalar  superpartner of 
$\psi_z$, modifies the ground state of 
the model, thus  loosing its properties like $m_{3/2}\sim\,$TeV.
However, tuning the coefficients ($b_K, b_{CK}$)  of the Kahler connection 
anomalies or a more complicated hidden sector could avoid this issue. 
Finally, Appendix~A  presents details on deriving  the  $U(1)_R$ charges of
the fields, in the conformal compensator description, and Appendix~B
reviews the ``naive'' flat-space field theory results for anomalies.

\section{Constructing models with gauged R-symmetry}
\label{sec1}

In this section we discuss models with a $U(1)_R$ 
gauge symmetry in 4D N=1 supergravity
 \cite{Cremmer:1983yr,Chamseddine:1995gb,Castano:1995ci,Binetruy:2004hh,Lee:2008zs}
(for the Lagrangian see \cite{WB,AP}).
The goal is to understand better   SUSY breaking, metastability 
of the ground state and to see if $m_{3/2}$ remains tunable to TeV values
in such models.
Consider a Kahler potential  $\cK$ and superpotential $W$ and define
\bea\label{gg}
G&=& \kappa^2 \cK +\ln \big[\kappa^6 \vert W\vert^2 \big]
\eea
Then the scalar potential of the model is $V=V_F+V_D$ with
\bea
V_D&=&\frac{1}{2} \,(\Re f(s))^{-1\, ab}\,D_a\,D_b
\nonumber\\
V_F&=&\kappa^{-4} \,e^G\,\Big[ G^i\,(G^{-1})^k_i\,G_j-3\Big] 
\nonumber\\[4pt]
&=& e^{\kappa^2 \cK}\,\Big[ (W^i+\kappa^2\,\cK^i\,W^\dagger) (\cK^{-1})^j_i
(W_j+\kappa^2\,\cK_j\,W)
-3\,\kappa^2\,\vert W\vert^2\Big]
\label{pot}
\eea
while the auxiliary fields $F^i$ 
\medskip
\bea\label{fpot}
F^i=- \kappa^{-1}\,e^{G/2}\,(G^{-1})^i_j\,G^j=
-e^{\kappa^2\,\cK/2} \, \frac{\vert W\vert}{W^\dagger}\,(\cK^{-1})^i_j\,
\big(W^j+\kappa^2\,\cK^j\,W^\dagger\big)
\eea

\medskip\noindent
in a standard notation\footnote{We use:
 $G^i=\partial G/\partial \phi_i^\dagger$, $G_j=\partial G/\partial\phi^j$,
$G^i_j=\partial^2G/\partial \phi^\dagger_i\partial\phi^j$, $G^i_j=\cK^i_j$.
$\cK_j=\partial \cK/\partial\phi^j$, $\cK^j=\partial \cK/\partial\phi_j^\dagger$, 
$W_i=\partial W/\partial\phi^i$ and $W^i=\partial W^\dagger/\partial\phi_i^\dagger$;
the index $i$ in $\phi^i$ labels all fields of the model, including the dilaton  $s$.
}; $f$ is the gauge kinetic 
function, $D_a$ are the Killing potentials \cite{WB,AP}
\bea\label{FF}
D_a\equiv -i \,\cF_a + i\, X_a^j(\phi)\, \cK_j,\qquad
\cF_a=-\frac{X_a^j\,W_j}{\kappa^2\,W},
\qquad
\Ra
\qquad
D_a=\frac{i\,X_a^j}{\kappa^2 \,W}\,\Big[W_j+\kappa^2\,\cK_j \,W\Big]
\label{dd}
\eea

\medskip\noindent
where $X_a^j$ are
 Killing vectors, with    
$\delta \phi^j=\Lambda^a\,X_a^j (\phi)$ under a gauge transformation
of parameter $\Lambda^a$. Also, the (gauge) covariant derivative is 
defined by\footnote{
To establish our conventions, 
for an Abelian case we use  $X^j\equiv -i\,q^j \phi^j$, therefore 
 $\phi^j\ra \exp(-i\,q^j\,\Lambda)\phi^j$ for a field $\phi^j$ of charge
$q^j$ under a gauge transformation 
$A_\mu \ra A_\mu +\partial_\mu \Lambda^a$, therefore
 $D_\mu\phi^j=(\partial_\mu \phi^j+i q^j A_\mu^a)\phi^j$.}
 $D_\mu \phi^j=\partial_\mu \phi^j-X_a^j A_\mu$.


\subsection{The model: 
de Sitter  ground state and TeV-gravitino mass from $U(1)_R$} 
\label{sec:vacuum}

In \cite{V1,IA} (also \cite{V2}) a class of metastable de Sitter vacua was 
discussed,
which have a tunable (infinitesimally small) value of the cosmological
constant and a TeV gravitino mass, based on a gauged $U(1)_R$.
We briefly review this model  in this section.
The spectrum consists of a  chiral multiplet (dilaton 
$S$) and a gauge multiplet of $U(1)_R$ in addition to the supergravity multiplet 
$(e_\mu^i,\psi_{3/2}$).
The chiral multiplet $S$ (dilaton) has  a shift symmetry ($\cgs$ is a real constant)
\bea
S \ra S - i\cgs \Lambda.
\eea
which is gauged. $\cK$ is defined below while
gauge invariance dictates the form of $W$
\medskip
\begin{align}
\mathcal K = -2 \kappa^{-2} \ln(s + \bar s),
 & & W= 
\kappa^{-3}  a\, e^{bs},
 \label{oldmodel}
\end{align}
where $a, b$ are dimensionless constants. In  string theory $b>0$
is considered  nonphysical and $b<0$ corresponds to a 
nonperturbative superpotential. We assume $b<0$.
The gauge kinetic function of $U(1)_R$ is  taken to be $f(S) = S$;  the dilaton transforms
non-linearly, then $X_a^s$ is field-independent, $X_a^s= - i \cgs$, see eq.(\ref{FF}).
 The scalar potential
is then, using eqs.(\ref{gg}) to (\ref{dd})
\medskip
\bea
V& =&   V_F +  V_D
\nonumber\\[3pt]
V_D &=&   \frac{\kappa^{-4}}{s + \bar s} \left[b\,\cgs
- \frac{2\,\cgs}{s+\bar s}\right]^2
\nonumber\\[3pt]
 V_F& =&  \kappa^ {-4}\, \vert a\vert^2\, e^{b(s+\bar s)} 
\left[  \frac{b^2}{2}  - \frac{2b}{s+\bar s}  - \frac{1}{(s+ \bar s)^2 }    \right] 
 \label{oldpotential}
 \eea

\medskip\noindent
Also $\cF_a= i\,b\,\cgs\,\kappa^{-2}\equiv - i\,\xi\,\kappa^{-2}$
is  the Fayet-Iliopoulos term. The auxiliary  $F^s$ of $S$ is 
\medskip
\bea
 F^s = - \frac{\kappa^2}{2} \, 
e^{\kappa^2 \mathcal K/2}  (s + \bar s)^2 \vert W\vert\,
\Big[ b - \frac{2}{(s + \bar s)}  \Big] 
= 
- \frac{  \kappa^{-1}  \vert a\vert   }{2}\,
e^{b\,(s+\overline s)/2}\, \left[ b \,(s + \bar s) - 2 \right]\
\label{Fs}
 \eea

\medskip\noindent
Minimising the scalar potential and imposing a  small  positive value for  
it at the minimum point,   $V_{\rm min}=\kappa^{-4}\epsilon_0$ where 
$\kappa^{-4}\epsilon_0\approx (10^{-3} eV)^4$,
 give
\medskip
\bea
\label{bsalpha}
\frac{e^{-\alpha}}{\alpha}
\frac{2 \,(\alpha-6)}{\alpha^2-2 \alpha  -2}=\frac{\vert a\vert^2}{b\,\cgs^2}
\eea
and respectively
\bea
\label{vmin}
\frac{e^{-\alpha}}{\alpha}\,\frac{(- 2) (\alpha-2)^2}{(\alpha-2)^2-6}
+\frac{2\,\epsilon_0\,e^{-\alpha} \alpha^2 b^{-3}}{[ (\alpha-2)^2-6]\, \cgs^2}
=\frac{\vert a\vert^2}{b\,\cgs^2},
\qquad
\alpha\equiv b \langle s+\overline s\rangle
\eea

\medskip\noindent
The second term in the lhs of eq.(\ref{vmin}) can be neglected to a good
approximation, so the existence of a minimum and a vanishing  $V_{\rm min}$ 
($\epsilon_0=0$) can both be fixed by one constraint on the set $\{a,b,\cgs\}$:
\bea\label{kz}
\alpha\approx -0.1833,\qquad \frac{\vert a\vert^2}{b\,\cgs^2}\approx -50.6557
\eea

\noindent
Note that $\alpha$ fixes only the product 
$b\,\langle \Re[s]\rangle$,
so any change of $b$ can in principle\footnote{This has
an impact on the value of $1/g^2_s\equiv \Re[s]=\alpha/(2\,b)$ where $g_s$ is the
4D  coupling.}
be compensated by that of $\Re[s]$.
A very small change of $\cgs$ can adjust a small (positive) cosmological constant
without impact on solution (\ref{kz}).
The spectrum  contains a   gravitino of  mass
\medskip
\bea
\label{m32}
\vert \,m_{3/2}\,\vert=\vert\,\kappa^2\,e^{\kappa^2\,\cK/2}\,W\,\vert
= 
\kappa^{-1} \Big\vert \frac{a\,b}{\alpha}\Big\vert \,\, e^{\alpha/2} 
\sim 1\,{\rm TeV}
\quad {\rm if}\quad \vert a\,b\vert\approx  200\,\kappa \times (1\, {\rm GeV})
\eea

\medskip\noindent
As shown above,  a TeV gravitino mass  is possible
if a second tuning constraint is satisfied
$\vert a\,b\vert\sim 8.3\times 10^{-17}$, by tuning $a$. 
For example,  $a\sim \cgs\sim 10^{-13}$ with $b\sim \cO(10^{-3}-10^{-2})$.

The scale of SUSY breaking is found after rescaling 
$s\ra s\kappa^{-1}$ and  $F^s\ra F^s\kappa^{-1}$ 
to restore their mass dimension (before this, $[s]\!=\!0$, $[F^s]\!=\!1$).
This gives  $\langle F^s\rangle\! \sim \! m_{3/2}\,m_P/{\vert b\vert}$
or
$\sqrt {\langle F^s\rangle}\! \sim\! 10^{10}/\sqrt{\vert b\vert}$ GeV, so we have gravity mediation,
with the mentioned values of $b$.

The mass of the dilaton $s$ 
 is found
\medskip
\bea
m_s^2=\frac{(s\!+\!\overline s)^2}{4}\langle \partial_s\partial_s V\rangle
=\frac{1}{4}
\frac{\kappa^{-2}\,b^3\,\cgs^2\,g(\alpha)}{\alpha^3\,(\alpha^2 -2 \alpha -2)},
\,\,
g(\alpha)\equiv\alpha^5-8\alpha^4+8\alpha^3+44 \alpha^2 -72 \alpha-24.
\eea

\medskip\noindent
Under the aforementioned constraints for parameters $a,b,\cgs$,  the mass of the dilaton
is of order $\cO(1-100)$ TeV 
\bea
m_s\approx 15.42 \vert b\vert^{3/2}\,\,\cgs\, \kappa^{-1}.
\eea
Finally,  the gauge field $C_\mu$ 
of $U(1)_R$ ``absorbs'' the axion ($\Im[s]$) and its mass
is found after canonical  normalisation of the gauge kinetic terms\footnote{
The ``unusual''  power of $3$ in $m_{C_\mu}$ accounts for the fact that the mass
is expressed in terms of Planck units and is consistent with the 
heterotic string result (where the power is 2) if
 expressed in terms of string scale.}
\bea
m_{C_\mu}= \frac{2\,\cgs \,\kappa^{-1}}{\langle s+\overline s\rangle^{3/2}}
\approx 25.48\times \vert b\vert^{3/2}\,\cgs\,\kappa^{-1}
\eea

\medskip\noindent
which is above the TeV scale for considered $a, b,\cgs$.

To conclude by tuning  $a, b, \cgs$ one can  ensure the existence
of a  ground state  with a small positive cosmological constant,
a  TeV-mass of the gravitino and a scale of SUSY breaking 
similar to that in gravity mediation.
These results for  a toy model  can be used to construct a more 
realistic model for SUSY breaking in supergravity with gauged $U(1)_R$;
however,   parameters $a, b,\cgs$ must then  satisfy 
extra constraints like  anomaly cancellation constraining $\cgs$ (see later),
with strong impact on the viability of the models (such as the values of the gravitino
and soft terms masses).

\subsubsection{The metastability of the ground state}

The ground state we found is metastable, since there is another minimum
$V_{\rm min}=0$ for
a runaway solution at $\Re[s]=\infty$.  We thus need to estimate
the probability ($\Gamma$) for the current vacuum  $V_{\rm min}=\kappa^{-4}\epsilon_0>0$ 
to decay into the vacuum $V_{\rm min}=0$ along the $\Re[s]$ direction,
 through the potential barrier (fig.1 in \cite{V1,IA}).
This is to ensure the ground state is lived long-enough. This probability is
(per unit of time and volume) 
\begin{align}
{\Gamma}&= \cA \, e^{-\frac{\cB}{\hbar}} \left(1 + O(\hbar) \right),
\end{align}
where $\hbar$ is the reduced Planck constant (we set $\hbar = 1$), 
and $\cA$ and $\cB$ depend on the model. 
The value of $\cA$ plays a minor role in comparison with the 
exponential suppression;
$\cB$ is fixed
 by the Euclidean action  of the instanton (bounce) solution ($S_1$)
which in the limit of a very small energy difference  between the two 
minima is \cite{Coleman:1977th}
\begin{align}
\cB  = \frac{27 \pi^2 S_1^4}{2\,\epsilon^3_0},
\label{B}
\end{align}
We redefine the field $s$ into (for standard kinetic terms; $s$ is dimensionless)
\begin{align}
\phi_s = \kappa^{-1} \log (s + \bar s),
\end{align}
$S_1$ is given by \cite{Coleman:1977th}
\begin{align}
S_1 &= \int d \phi_s \sqrt{2 \cV(\phi_s)}=\kappa^{-3} \vert a\vert \,\cS(b)
\end{align}
where  $\cV$ is the potential of $\phi_s$ and
\medskip
\bea
\cS(b)& =& \sqrt 2 \int_{\ln\left( \frac{\alpha}{b} \right)}^\infty d x 
\,  
\Big[e^{b\, e^{x}} \left( \frac{b^2}{2} -2\, b e^{-x} - e^{-2x} \right)
+  
\frac{e^{-x}}{b\, \zeta(\alpha)} \left( 2\, e^{-x} - b\right)^2  \Big]^{1/2}
\eea

\medskip\noindent
which can be computed numerically. 
$\zeta(\alpha)$ is given by the expression in the  lhs of eq.(\ref{vmin})
(where one can set $\epsilon_0=0$ to a good approximation). 
If we  interpret $\Re(s)$ as the (inverse of the) 
4D coupling ($1/g_s^2$) to a GUT-like value,
 $\Re(s)\approx 25$, this  fixes $b$ via
$b=\alpha/\langle s+\overline s\rangle\approx -3\times 10^{-3}$. 
For this value of $b$ one finds  $\cS \approx 0.0124106$.
By demanding that the gravitino mass be of order 
TeV (by tuning  $a$, see previous section) 
one finds  $ \cB \approx 10^{297}$. Therefore $\Gamma$ is extremely  small
(largely due to the small difference between the two minima)\footnote{
Usually values of $\cB\geq 400$ are regarded as metastable enough \cite{Blum:2009na}.};
the ground state  is long lived enough to use 
this model as a starting point for building  realistic models, 
by adding  physical fields, that have such ground state {\it along}
this field direction.

\subsection{A toy model with an additional field in the visible sector} 
\label{toy_model}
\label{section2.2}

We would like to preserve the nice features of the previous 
toy model and its ground state, while  coupling the model to the visible sector. 
We thus add to the model  one physical scalar  field  ($\phi$) 
not charged\footnote{Its fermionic spartner $\psi$ has however a  $U(1)_R$
 charge $R_\psi=R_\phi+\xi/2=\xi/2$, see the Appendix.} under $U(1)_R$. 
A general way to couple the visible to the hidden sector 
while preserving the gauged $U(1)_R$ is to consider
\begin{align}
\cK=-2\,\kappa^{-2}\,\ln (s+\overline s)+\phi^\dagger\phi,\qquad\quad
W = A(\phi) \,e^{bs} + B(\phi). 
\label{W_phys}
\end{align}

\medskip\noindent
 $W$ generalises the form in eq.(\ref{oldmodel}).
If $B(\phi) \neq 0$, this implies that the scalar field $\phi$ in the visible sector
should have an R-charge, which we do not consider in this sub-section. 
So $B=0$ and
\begin{align}
 W = A(\phi) \,\, e^{bs}.
\end{align}
With this $W$, the
 scalar potential is then $V = V_F +  V_D$, see eq.(\ref{pot}), where
\medskip
\bea\notag 
 V_D &= & \frac{\kappa^{-4}}{s + \bar s} \left[
b\,\cgs- \frac{2\,\cgs}{s+\bar s}\right]^2    , 
\qquad
\sigma_s=\frac{1}{2} \,\big[b\,(s+\overline s)-2\big]^2-3
\nonumber\\[4pt]
 V_F &=&
 \frac{e^{\kappa^2 \phi^\dagger \phi} \,  }{(s + \bar s)^2} \left[\,
\kappa^2\,\sigma_s\, |W(\phi)|^2 +  |\nabla_\phi W (\phi)|^2 \,\right] 
\label{toy_potential}
\eea

\medskip\noindent
and  $\nabla_\phi W =  (\partial /\partial\phi + \kappa^2  \phi^\dagger) W$.
The auxiliary fields $F^s$, $F^\phi$ of $s$ and $\phi$ are 
\bea\label{kk}
F^s &=& -e^{\kappa^2 \,\cK/2}\,\vert W\vert \,\frac{(s+\overline s)^2}{2}\,\kappa^2\,
\Big(b-\frac{2}{s+\overline s}\Big)
\nonumber\\
F^\phi &=& - e^{\kappa^2\,\cK/2}\,\frac{\vert W\vert}{W^\dagger}
\,\,\big[\nabla_\phi W \big]^\dagger
\eea

\medskip
For phenomenological reasons, in viable models one
would like to avoid SUSY breaking by the ``visible'' sector, so
we  demand $\langle F^\phi\rangle =0$ or 
 \begin{align}
 \langle \nabla_\phi A(\phi)\rangle  = 0. 
\label{noSUSYbreaking}
 \end{align}
With $A(\phi)$ being analytical, one
sees that the only possibility to respect this is to have 
$\langle \phi\rangle = 0$, $\langle \partial_\phi A\rangle=0$.
This gives 
\begin{align}
\langle A(\phi) \rangle = \kappa^{-3} a.
\end{align}
Therefore
\medskip
\bea\label{W}
W=\big[\kappa^{-3}\,a+\tilde W(\phi)\big]\,e^{b\,s},
 \quad \langle \tilde W(\phi)\rangle =0.
\eea

\medskip\noindent
The ground state of the model is found by minimising $V$ wrt $\phi$ and $s$ 
\bea
\frac{\partial V}{\partial\phi}=0,\qquad
\frac{\partial V}{\partial s}=0
\eea
The former is automatically respected with eq.(\ref{noSUSYbreaking}), also given that
$\phi$ is not R-charged. 
The second condition gives
\medskip
\bea
\label{min2}
e^{\kappa^2 \langle \vert \phi\vert^2\rangle+
b\,\langle s+\overline s\rangle}\,
\langle \rho_s\rangle\, 
\vert a\vert^2
=
\cgs^2 \,\Big(b-\frac{6}{s+\overline s}\Big),
\eea
where
\bea
\rho_s=\frac12\,\big[ b^2 (s+\bar s)^2- 2 b (s+\bar s) -2\big].
\eea

\medskip\noindent
which is  identical to  (\ref{bsalpha}) for $\langle \phi\rangle=0$.
Thus the previous ground state in eq.(\ref{bsalpha}) defined by
$\langle s+\overline s\rangle=\alpha/b$ 
  is preserved, with $\langle\phi\rangle=0$.

Further, using the notation $w_{\phi\phi}\equiv\partial^2 \tilde W(\phi)/\partial\phi^2$,
we find
\medskip
\bea
\langle \partial_\phi \partial_\phi V \rangle
&=&\frac{ e^{\alpha} }{\alpha^2} 
\,\, b^2\, \big(\langle \sigma_s\rangle+2\big) \,\,a^\dagger\,\kappa^{-1}\,\langle
w_{\phi \phi}\rangle
\notag
\\[4pt]
\langle \partial_\phi \partial^{\bar \phi} V\rangle &=& 
\frac{ e^\alpha }{\alpha^2}
\,b^2\,\big[\,\big(\langle \sigma_s\rangle+1\big)\,\vert a\vert^2\,\kappa^{-2}\,
+ 
\vert\langle w_{\phi\phi}\rangle \vert^2 
\big]
\label{scalarmass}
\eea

\medskip\noindent
while  $\langle \partial_\phi\partial_s V\rangle=0$
following from  $\langle F^\phi\rangle=0$.
The  eigenvalues of the above mass matrix give
\begin{align}
m^2_\pm = \frac{e^\alpha}{\alpha^2} 
\,b^2\,\,\Big[\,
 \kappa^{-2} \,\,
a^2\, (\langle \sigma_s\rangle +1)  + |w_{\phi \phi}|^2 \pm (\langle \sigma_s\rangle
 +2 ) \,
\kappa^{-1}\, | a^\dagger w_{\phi \phi} | \,
\Big]
\end{align}
If $ \kappa^{-1} \vert a\vert\gg \vert w_{\phi \phi} \vert $, 
$m_\pm\sim m_{3/2}$ where $m_{3/2}$
remains equal to that found previously in eq.(\ref{m32}).
Finally the mass of the fermion $\psi$, superpartner
 of $\phi$, is
$m_{\psi} = w_{\phi \phi} e^{b \langle s \rangle}/(\langle s + \bar s \rangle)$.

In conclusion, the guideline to construct  a realistic model is this:
replace the visible sector field  $\phi$  by the MSSM  superfields assumed 
to be $U(1)_R$ neutral, just like the field $\phi$.
Then  the ground state with all previous benefits (TeV gravitino mass, gravity mediated 
SUSY breaking etc) is preserved\footnote{Such a case is discussed in Section~\ref{sec4}.}.
Then from eq.(\ref{W}) we conclude  that
 a minimal realistic model has
\medskip
\bea\label{mssm}
W=\big[\kappa^{-3}\,a+\tilde W_{MSSM}\big]\,e^{b\,s}
\eea

\medskip\noindent
where $\tilde W_{MSSM}$ is the usual  MSSM superpotential with
superfields replaced by their scalar components. 
The Kahler potential will be similar to that considered here, with
the contribution of $\phi$ replaced by that of the MSSM fields.
The squared soft scalar masses  of such model can be shown to be positive
and close to the square of the gravitino mass (TeV$^2$).

\bigskip
\subsection{A toy model with a  $U(1)_R$-charged field in the hidden sector}
\label{section2.3}

The model considered in the previous section is still too simple in 
that the hidden sector contains only the
scalar  field $s$ and a $U(1)_R$ gauge boson and their superpartners. 
Ensuring the quantum consistency of the model (anomaly cancellation condition\footnote{This is 
discussed in Section~\ref{anomalies}.}) and maintaining at the same time the nice
properties of the ground state (TeV-scale gravitino, etc) 
  may demand the presence of additional 
$U(1)_R$ charged fields in the hidden sector.

Let us then add an extra scalar field $z$ (and its superpartner)
in the hidden sector, with a  $U(1)_R$ charge $q_z$,  
and examine when the  ground state of the initial model  can be  maintained,
together with its benefits. The visible sector scalar field  
$\phi$ remains $U(1)_R$-neutral (while its superpartner acquires a $U(1)_R$ charge).
As a further step from eq.(\ref{W_phys}), we take
\medskip
\begin{align}
\cK=-2\,\kappa^{-2}\,
\ln (s+\overline s)+\phi^\dagger\phi+z^\dagger z,\qquad\quad
 W = A(\phi,z) \,e^{b\,s}. 
\label{w3}
\end{align}

\medskip\noindent
where we omitted a possible addition of a function $B(z)$ in $W$ 
for simplicity and for preserving the basic properties of the model.
From the invariance of the action under $U(1)_R$, we find the fields 
transformations  under $U(1)_R$
\medskip
\bea\label{op}
s &\ra& s - i \,\cgs\Lambda, 
\qquad\qquad
A(\phi,z) \ra  A(\phi,z)\,e^{-i\,\beta\,\Lambda},
\nonumber\\
z &\ra& z \,e^{-i q_z\,\Lambda},
\qquad\qquad\,\,\,
W(\phi,z) \ra W(\phi,z)\,e^{-i \,(\cgs\,b+\beta)\,\Lambda}
\eea

\smallskip\noindent
so that $\delta A=-i\,\beta 
\,A\,\Lambda$ and with $\delta A=\partial_z A \,\delta z=-i\,q_z\,z\,\Lambda
\,\partial_z A$ then  $\partial\ln A/\partial\ln z=\beta/q_z$. Thus
$A(\phi,z)=(\kappa\,z)^{\beta/q_z}\,
[a\kappa^{-3}+ \tilde W(\phi)]$, where  $a$ is a constant
and $W=(\kappa\,z)^{\beta/q_z}\,
\big[a\,\kappa^{-3}+ \tilde W(\phi)\big]\,e^{b\,s}
$.
The scalar potential is then $V=V_F+V_D$ with
\medskip
\bea\label{uf}
V_D&=&\frac{\kappa^{-4}}{s+\overline s}
\, \Big[\beta+\kappa^2\,q_z \vert z\vert^2
+b\,\cgs - \frac{2\,\cgs}{s+\overline s}\Big]^2
 \nonumber\\[5pt]
V_F
&=&
e^{\kappa^2\,\cK}\,\,
\Big[
\kappa^2\,\sigma_s\,\vert W\vert^2 
+\vert W_\phi+\kappa^2\,\phi^\dagger  \,W\vert^2
+\vert W_z+\kappa^2\,z^\dagger  \,W\vert^2
\Big],
\eea

\medskip\noindent
The Fayet-Iliopoulos 
term is now $\cF_a=i\, (\beta+ b\,c_{GS})\,\kappa^{-2}\equiv -i\,\xi\,\kappa^{-2}$.
Again we demand 
that\footnote{$F^s$ and $F^\phi$ are those of (\ref{kk}) while $F^z=F^\phi\vert_{\phi\ra z}$.}
 $\langle F^\phi\rangle=0$ or $\langle \nabla_\phi W\rangle=0$
 which is satisfied for a standard $\tilde W$, polynomial in fields
 (such as that  of the MSSM)  and $\langle \phi\rangle=0$.

The minimum condition of the potential wrt $\phi$ is automatically satisfied 
since $\phi$ does not
break SUSY.
To simplify the analysis we consider that 
the scalar field  $z$ does not enter in the
superpotential, so $\beta=0$. This choice is also 
motivated by the analysis in the next section
and impacts on the value of FI term $\xi$, giving
$\xi=-b\,\cgs$ 
as we had before\footnote{This is important since $\xi$ is related to the
R-charges of the fields and plays a role in anomaly cancellation.}.
Then
\medskip
\bea
W=\big[ a\kappa^{-3}+\tilde W(\phi)\big]\,e^{b\,s}
\eea

\medskip\noindent
The minimum conditions, evaluated at $\langle\phi\rangle=0$ give
\medskip
\bea\label{m1}
e^{\gamma+\alpha}\,b\,\vert a\vert^2\,\big(\langle\rho_s\rangle+\gamma)(1-2/\alpha\big)
-
\big[ b\cgs\,(1-2/\alpha)+\gamma\,q_z\big]\,
\big[ b\,\cgs\,(1-6/\alpha)+\gamma\,q_z\big]&=&0
\\[6pt]
\Big[e^{\gamma+\alpha}\,(b/\alpha)\,\vert a \vert^2\,\big(\langle \sigma_s\rangle +1+\gamma\big)
\,+
2\,q_z\,
\big[q_z\,\gamma+b\,\cgs (1-2/\alpha\big]\Big]\langle z^\dagger\rangle \,(b/\alpha)&=&0
\qquad
\label{m2}
\eea

\medskip\noindent
where $\gamma=\kappa^2\,\vert \langle z\rangle \vert^2$ and
with $\alpha=b\langle s+\bar s\rangle$,
 $\langle \sigma_s\rangle=(1/2)(\alpha-2)^2-3$, 
$\langle \rho_s\rangle=(1/2) (\alpha^2-2 \alpha-2)$.
 The condition $V_{\rm min}=\kappa^{-4}\epsilon_0=(10^{-3} {\rm eV})^4$ gives
\medskip
\bea\label{m3}
e^{\alpha+\gamma} \,\vert a\vert^2 (b/\alpha)\,
\big(\langle \sigma_s\rangle+\gamma\big)
+\big[ q_z\,\gamma+b\,\cgs \,(1-2/\alpha)\big]^2=\epsilon_0\,\alpha/b.
\eea
 
\medskip\noindent
The system  of eqs.(\ref{m1}), (\ref{m2}) should be solved for $\alpha$ and 
$\gamma$ in terms of the parameters of the model. 
One can see that $\langle z\rangle =0$ is a solution to (\ref{m2})
which if used in eqs.(\ref{m1}), (\ref{m3}) gives
two equations identical to (\ref{bsalpha}), (\ref{vmin}) and thus
have the same solution $\alpha=-0.1833$ as found there. With this value of 
$\alpha$, the square bracket in eq.(\ref{m2}) is positive and nonzero if $q_z<0$ and
$\cgs>0$.  In this situation,  the original ground state 
$\langle s+\overline s\rangle =\alpha/b$, ($b<0$)
together with $\langle \phi\rangle=\langle z\rangle=0$ is indeed a solution
of this extended model provided that
 $q_z<0$, $\cgs>0$.  If these conditions are violated, then $\langle z\rangle\not=0$ 
and the original vev for $s$ and its properties are not maintained.
It can be shown that this solution is a  ground state (i.e. local minimum) if
 a  higher order Kahler term with $(\partial \cK/\partial z)_o=0$ is added to $\cK$,
without modifying this solution.

In conclusion,  a hidden sector field with charge $q_z<0$ can be added to the
model of previous section without altering the vacuum of the initial theory, 
with the new field directions having vanishing vev's. As a result, 
$m_{3/2}$  is the same as found before (independent of  $q_z$). 
Anomaly cancellation to  which the superpartner 
fermion of  $z$  will contribute can constrain $q_z$.
We shall then check if  $q_z<0$, $\cgs>0$ are
consistent with anomaly cancellation (Section~\ref{sec4}).

\section{$U(1)_R$ anomalies in supergravity  and their field theory view}

\label{anomalies}

The discussion so far was at the classical level.
Quantum consistency  like  anomalies cancellation is a strong
constraint, discussed next. This can change the results we found,
such as the TeV-values for $m_{3/2}$, because the  parameters
of the models ($a,b,\cgs$ or charges) are more constrained now.
This section is also relevant on its own, independent of the 
rest of the paper.

Anomaly cancellation   in supergravity with an anomalous $U(1)$ 
with Fayet-Iliopoulos term(s) and Green-Schwarz mechanism 
where discussed in the past  in  a general setting in \cite{F1,F2}  
that we use below. However, the relation and agreement 
of such results to the ``naive'' field theory approach
was not examined for the rather special case of  a  gauged $U(1)_R$.
In fact in global SUSY $U(1)_R$ cannot even be gauged. 
In this section we carefully investigate this relation.  
We refer the reader to these two papers for the details of the supergravity 
analysis while in  Appendix~B we review the naive  field theory 
results (flat space), with our conventions for  the $U(1)_R$ charges 
defined in Appendix~A. For a previous detailed study see \cite{Chamseddine:1995gb}.

The field content is that of minimal supergravity, with gauged $U(1)_R$
and the dilaton $S$ for the GS mechanism, as in previous sections, and
with additional (say MSSM-like)  superfields and SM gauge group times $U(1)_R$.
Anomaly cancellation conditions  for the cubic anomaly $U(1)_R^3$  ((a) below)   
and the mixed anomalies of $U(1)_R$  with the Kahler connection 
$K_\mu$ ((b), (c)), with the SM gauge group (d) and  gravity (e) 
were found in eqs.(4.4) in Section~4 of \cite{F1}. 
These  are
\medskip
\bea\label{an}
(a)\qquad\,\,\quad
{\rm C\tilde C:}\qquad
 0 &=& \Tr\big[(Q+\xi/2)\,Q^2\big]
+\xi\,a_{KCC}- \cgs \,b_{C}
\nonumber\\[2pt]
(b)\qquad\quad C\tilde K: 
\qquad
0 &=& \Tr\big[(Q+\xi/2) Q\big]
-\xi\,a_{CKK}-a_{KCC} - \cgs \,b_{CK}
\nonumber\\[0pt]
(c)\quad \qquad K\tilde K: 
\qquad 0&=& \Tr\big[Q\big]
-\frac{1}{2}\, \xi\,(n_{\lambda}+3-n_\psi)
+4\,a_{CKK}-4\,\cgs \,b_{K}
\nonumber\\
(d)\quad \quad
(F\tilde F)_\alpha: \qquad
0&=& - \Tr\big[ Q\,(\tau^a \tau^b)_\alpha\big]+\frac{\xi}{2}\,\big[C_2(G)-C(r)\big]_\alpha\delta^{ab}
+\frac{1}{3} \,\cgs\,b_{A,\alpha}\,\delta^{ab}
\nonumber\\
(e)\,\,\quad
\qquad
\cR\tilde \cR: \qquad
0&=&
\Tr\big[Q\big]-\frac{\xi}{2}\,(n_\lambda-21-n_\psi)
+8\,\cgs\,b_{R}
\eea

\medskip\noindent
where $\,\tilde{}\,\,\,$  labels the dual field strength.
$C_\mu$ $(C_{\mu\nu})$ is the gauge field (strength) of $U(1)_R$;
$F_{\mu\nu}$ is a field strength  corresponding to gauge fields $A_\mu$ of
the SM gauge group (with group generators\footnote{Refs.\cite{F1,F2}
use anti-hermitian  $T^a\!=\!-i\tau^a$,
so we added a minus  in front of $\Tr$ in line (d) of eqs.(\ref{an}).} $\tau^a$); 
$\alpha$ is a group index that runs over $U(1)_Y$, $SU(2)_L$, $SU(3)$.  
$K_\mu$ is the Kahler connection that essentially fixes the coupling of 
the gravitino. These three fields are involved in
 conditions (a), (b), (c), (d).
Condition (e) is for the mixed, $U(1)_R$-gravitational anomaly.
There is also the usual SM condition 
$\Tr[\tau^a\{\tau^b,\tau^c\}]=0$ for SM group generators.

Regarding notation, 
$n_\lambda$ is the number of gauginos present, $n_\psi$  is the number of Weyl fermions
(matter fermions and dilatino),
$Q$'s are the $U(1)_R$ charges of matter superfields 
and $\xi$ is the Fayet Iliopoulos  term of our $U(1)_R$, (see also Appendix~A).
Unlike other types of anomalous $U(1)$'s, 
for the special  case of  $U(1)_R$,  the SM  gauginos, $U(1)_R$ gaugino 
 and gravitino are all {\it charged under $U(1)_R$}  and 
contribute to  the anomalies. 
Also $\Tr_r(\tau^a \tau^b)=C(r)\,\delta^{ab}$
is the trace over the irreducible representations $r$, 
and $f^{acd} f^{bcd} =\delta^{ab} C_2(G)$.
The terms in eqs.(\ref{an}) 
proportional to $\xi$ of coefficients $+3$ and $-21$
denote the contributions of the gravitino to those anomalies \cite{grisaru}.
Finally, $\cgs$ is the GS coefficient and $a_{KCC}$, $a_{CKK}$ 
are coefficients of local counterterms that can be present (defined shortly).

To understand the role of Kahler connection $K_\mu$ 
it is instructive to write down the gauge ($V_\mu$) couplings,
present, in addition to the spin-connection,  in the
covariant derivatives of the gravitino $\psi_{3/2}$, gauginos $\lambda$ and matter 
fermions $\psi$ (such as those in the MSSM)\footnote{Note that eqs.(\ref{fts})
apply  for the case of flat field-space.}:
\medskip
\bea\label{fts}
\psi_{3/2}: & \qquad & V_\mu=-(i/2) \,K_\mu
\nonumber\\
\lambda^a: & \qquad & V_\mu^{ab}=-A_\mu^c \,f^{abc} - (i/2)\,K_\mu \delta^{ab}
\nonumber\\
\psi^\alpha: & \qquad & V_\mu= A_\mu^a\,(-i\tau^a) + i\,Q\,C_\mu +(i/2)\,K_\mu
\eea

\medskip\noindent
The local counterterms that come with coefficients $a_{KCC}$, $a_{CKK}$
are \cite{F1,F2}
\medskip
\bea
\delta \cL_1=\frac{1}{24\pi^2}\,\epsilon^{\mu\nu\rho\sigma}\,
\Big[
a_{CKK}\,C_\mu\,K_\nu\,\partial_\rho K_\sigma
+
a_{KCC}\,K_\mu\,C_\nu \partial_\rho C_\sigma
\Big]
\eea

\medskip\noindent
and $b_C$, $b_{CK}$, $b_K$, etc, 
 of eqs.(\ref{an}) are coefficients present in the Chern-Simons terms
\medskip
\bea\label{zz}
\delta \cL_2\!\!\!\!&=&\!\!\!\!
\frac{1}{48\pi^2}\,\Im[s]\,\epsilon^{\mu\nu\rho\sigma}
\partial_\mu\Omega_{\nu\rho\sigma}
\\[6pt]
\!\!&=&
\!\! %
\frac{1}{48\pi^2}\,\Im[s]
\Big[
b_{C}\,C_{\mu\nu}\,\tilde C^{\mu\nu}\!
+b_{CK}\,C_{\mu\nu}\,\tilde K^{\mu\nu}\!
+b_{K}\,K_{\mu\nu}\,\tilde K^{\mu\nu}\!
+b_{A,\alpha}\,(F_{\mu\nu}\,\tilde F^{\mu\nu})_\alpha\!
+b_{R}\,\cR_{\mu\nu}\,\tilde \cR^{\mu\nu}
\Big]
\nonumber
\eea

\medskip\noindent
with dual field definition $\tilde C^{\mu\nu}=1/2\,\epsilon^{\mu\nu\rho\sigma} C_{\rho\sigma}$, etc.
Some of  the coefficients
$b_C$, $b_K$, $b_{CK}$, $b_R$ can be related, 
as for example in heterotic
string theory (for $b_C, b_R$) and cannot  be adjusted at will. 
However, this applies only to anomalous $U(1)$'s which are not of $R$-type,
since it is difficult to derive  $U(1)_R$  from  strings. We 
 relax this constraint and consider them independent.

It is assumed that the dilaton ($S$)
is the only (super)field that transforms non-linearly under the gauged $U(1)_R$ and thus it
implements  the Green-Schwarz mechanism. 
The canonical normalization of the gauge kinetic term
for $U(1)_R$ gauge field ($C_\mu$) gives 
$b_C=12\pi^2$
 and we  assume that this is the case in the following,
while $b_K$ and $b_{CK}$ can in general be non-zero.

Supergravity conditions (\ref{an}) are not transparent from the ``naive''
field theory point of view for the anomalies of $U(1)_R$.
So let us clarify the link of these conditions to
 the  field theory result  in Appendix~B.
First, the $U(1)_R$ charges of the fields 
are shown below (see Appendix~A) and depend on the FI term(s):
\bea
R_\psi=Q+\xi/2,\qquad R_\lambda=R_{\psi_{3/2}}=-\xi/2,\qquad R_{\psi_s}=\xi/2.
\eea
where $Q$ is the charge of the superfield or scalar superpartner of $\psi$;  
$R_\lambda$, $R_{\psi_{3/2}}$ and $R_{\psi_s}$ are
 the charges of the gaugino $\lambda$, gravitino $\psi_{3/2}$ and  dilatino $\psi_s$, 
respectively.
Using this information, the first three relations in eqs.(\ref{an}) can be combined, 
after multiplying them by $4$, $4\xi$ and $\xi^2$ respectively and then 
adding them and using that $\Tr 1=n_\psi-1$. The result is\footnote{ 
``-1'' in $\Tr 1\!=\! n_\psi\!-\!1$  isolates 
the dilatino from matter fermions $\psi$ (the charge $Q$ of the
dilaton is 0.).}
\bea\label{fieldtheory}
\Tr \,[R_\psi^3]+n_\lambda\,R_\lambda^3 \,+3\,R_{\psi_{3/2}}^3+R_{\psi_s}^3
=  \cgs\Big[ b_{C}+\xi\,b_{CK} +\xi^2 \,b_{K}\Big]
\eea

\medskip\noindent
One recognizes 
in the lhs the usual field theory cubic $U(1)_R^3$ anomaly cancellation  condition 
in the presence of FI terms and GS mechanism, in which all fermionic
contributions are added and compensated by a  GS shift in the rhs:
the trace adds all matter fermions contributions,
$n_\lambda$ is the number of gauginos\footnote{For the $U(1)_R\times$SM
gauge group, $n_\lambda=1\!+\!(8\!+\!3\!+\!1)\!=\!13$,  since the $R-$gaugino contributes, via
 $\overline\lambda\,\lambda\,V_{\mu, R}$.}
each of a contribution $R_\lambda^3$;
$(+3) R_{\psi_{3/2}}^3$ is the gravitino contribution, 
three times larger than that of one gaugino \cite{grisaru}.
The result above has (with $b_C\!=\!12\pi^2$ and $b_{CK}\!=\!b_{K}\!=\!0$) the same form as 
 the ``naive'' field theory result, eqs.(\ref{an1}),(\ref{an2})
in Appendix~B.
This is interesting since in global SUSY $U(1)_R$ cannot even be gauged.
Note however the  difference in the rhs  due to 
  $\xi b_{CK}+\xi^2 b_K$. The terms in $\delta\cL_2$ of coefficients
$b_K$, $b_{CK}$ are not present in naive field theory case, 
and give extra freedom in canceling this anomaly.

The two remaining independent conditions of constraints (a), (b),  (c)  in eq.(\ref{an}),
refer to Kahler and mixed $U(1)_R$-Kahler connection. They  
can always be respected by a suitable choice of $a_{KCC}$ and $a_{CKK}$
of the local counterterms shown and  are not discussed further.
One finds (by  combining these two remaining constraints with  
condition (e) in eq.(\ref{an})), that  $a_{CKK}=3\xi\! +\! \cgs (b_K\!+\!2 b_R)$, while 
$a_{KCC}$ is found from one of equations (a), (b), (c) in eq.(\ref{an}).

The last two  conditions in eq.(\ref{an}) can be re-written as 
\medskip
\bea\label{ee}
(F\tilde F)_\alpha:
\quad\qquad\quad
\Tr\,\big[R_\psi\,(\tau^a\tau^b)_\alpha\big]+C_2(G_\alpha)\,\delta^{ab}\,R_\lambda
&=& (1/3) \,\cgs\,b_{A,\alpha}\,\delta^{ab}
\nonumber\\[9pt]
\cR\tilde \cR:
\qquad
\Tr\,\big[R_\psi\big]+n_\lambda\,R_{\lambda}+ (-21)\,R_{\psi_{3/2}}+R_{\psi_s}
&=& - 8\,\cgs\,b_{R}\qquad\qquad
\eea

\medskip\noindent
where  $C_2(G)\delta^{ab}=f^{acd} f^{bcd}$ with $C_2(G)=N$ for $SU(N)$  and 0 for
 $U(1)$, $\alpha$  labels the groups $U(1)_Y$, $SU(2)_L$, $SU(3)$.
The lhs of the first equation is exactly the naive
field theory contribution from (MSSM) matter fermions, of $R_\psi=Q+\xi/2$,
and  gaugino. In the second equation, there are contributions
 of: $n_\lambda$ gauginos of the SM {\it and} $U(1)_R$ gauge groups; 
   gravitino contribution  ($(-21)$ times 
that of a gaugino \cite{grisaru}), dilatino and 
the  $\Tr$ is over all  matter fermions.
These equations agree with the naive field theory result, eq.(\ref{an1}), (\ref{an2})
for corresponding anomalies.
$b_{A,\alpha}$ in the rhs of  the first equation  
also enters in the counterterm due to the GS mechanism in eq.(\ref{zz}).
By supersymmetry, it shows that the gauge kinetic function  coupling S to all SM sub-groups
becomes $k_{\alpha}\,S$ with  $k_\alpha\equiv b_{A,\alpha}/(12\pi^2)$ \footnote{
The quantity $b_{A,\alpha}/12\pi^2=k_\alpha$ plays the role of Kac-Moody levels in the heterotic string.}.
As for the rhs of the second condition in eq.(\ref{ee}), in field theory one has
$b_R=-12\pi^2$, eqs.(\ref{an1}), (\ref{an2}), while in supergravity one
is free to adjust this coefficient (unlike in heterotic string case).
This ends the relation of the anomaly cancellation conditions to the 
naive field theoretical results (global SUSY) obtained using
the $\Tr$ over the charged states.

\section{Application to MSSM-like models}
\label{sec4}

Let us now first apply these anomaly  relations to the model of Section~\ref{sec:vacuum}.
The $U(1)_R$-charged fermions of the model are the gravitino,
gaugino of $U(1)_R$ and dilatino.
Their contributions to the cubic $U(1)_R^3$  anomaly cancellation 
of eq.(\ref{fieldtheory})
give (see also the Appendix)
\bea\label{hu}
12\pi^2 \,\cgs= 
3\,(-\xi/2)^3
+ 
(-\xi/2)^3
+
(\xi/2)^3,\quad
\Ra 
\quad 
32\pi^2\,\cgs=-(-b\cgs)^3.
\eea

\medskip\noindent
since in such model $\xi=-b\,\cgs$.
This gives $b>0$, which contradicts our initial assumption $b<0$ needed
for the non-perturbative superpotential. 
Therefore this minimal model is inconsistent at the quantum level
and this demands the presence of extra states charged under $U(1)_R$. 

Let us then consider a  more  realistic model.
We  assume the presence of  the MSSM superfields in the visible sector
as outlined at the end of  Section~\ref{toy_model}.
For simplicity we assume  that  all MSSM
superfields have $U(1)_R$ charges equal to $Q$, which is
 indeed possible\footnote{Our $U(1)_R$ is realized only locally and 
we do not consider traditional R-parity symmetry as part of it \cite{Chamseddine:1995gb}}.
We also allow for the presence of an R-charged superfield $(z,\psi_z)$,
{\it singlet} under SM group, which  contributes only to anomalies 
that do not involve the SM group.

For the mixed anomalies of $U(1)_R$ with each of the subgroups
of the SM:  $U(1)_Y$, $SU(2)_L$, $SU(3)$,   we use eq.(\ref{ee}) with 
appropriate generators and obtain, respectively, the three 
equations below. 
Using the  quantum numbers of the states charged under the SM group:
$q:(1/6, 2,3)$,
$u^c:(-2/3, 1, \bar 3)$,
$d^c:(1/3, 1,\bar 3)$,
$l:(-1/2, 2. 1)$,
$e^c:(1,1,1)$,
$\tilde h_{1,2}:(\pm 1/2,2,1)$ then
 \medskip
\bea\label{sm}
11\,(Q+\xi/2)&=& (1/3)\,\cgs\,\, b_{A,1}
\nonumber\\[5pt]
7 \,(Q+\xi/2) - 2 \,\,(\xi/2)&=& (1/3)\,\,\cgs\,b_{A,2}
\nonumber\\[5pt]
6 \,(Q+\xi/2) - 3\, \,(\xi/2)&=& (1/3)\,\,\cgs\,b_{A,3}.
\eea

\medskip\noindent
To derive eqs.(\ref{sm}), we used:
$\Tr[(Q+\xi/2)\,Y^2_\psi]=(Q+\xi/2)\times 
3(1/2+1+1/6+4/3+1/3)+1/2(1+1)=11(Q+\xi/2)$ due, in order, to
$l$, $e^c$, $q$, $u^c$, $d^c$, $\tilde h_1$, $\tilde h_2$; for 
$SU(2)_L$:  $\Tr[(Q+\xi/2)\,T^a T^b]=
(Q+\xi/2) [\, 3(1/2+3/2)+1/2\,(1\!+\!1)]=7(Q\!+\!\xi/2)$, 
from $l$, $q$, $\tilde h_1$, $\tilde h_2$. Finally, for
$SU(3)$: $\Tr[(Q+\xi/2)\,T^a T^b]\!=\! (Q+\xi/2)\,\times 3\,(2\,\times 1/2+1/2+1/2)=
6\,(Q+\xi/2)$ from $q$, $d^c$, $u^c$. 

The above conditions 
become 
\medskip
\bea
\label{sm2}
\frac{11\,(Q+\xi/2)}{k_1}=
\frac{7\,Q+5\,(\xi/2)}{k_2}=
\frac{6\,Q+3\,(\xi/2)}{k_3}=
4\pi^2 \cgs,\qquad
k_\alpha\equiv \frac{b_{A,\alpha}}{12\pi^2}.
\eea

\medskip\noindent
where  $k_\alpha$ play the role
of Kac-Moody levels, see eq.(\ref{zz}) and Appendix~B
\footnote{The normalized kinetic terms are
$1/4\,\int d^2\theta\,\, k_\alpha\, S\, \Tr( W^a W_a)_\alpha$,
with $k_\alpha=b_{A,\alpha}/(12\pi^2)$ as Kac-Moody levels.}.
The cancellation of anomalies thus demands relation (\ref{sm2})
among the coefficients $b_{A,\alpha}$, $\alpha:1,2,3$, 
 which can be tuned to this purpose.
This relation is similar to eq.(\ref{an1}), (\ref{an2}).

By supersymmetry, the three gauge couplings of the SM group are
related to coefficients $b_{A,\alpha}$ and the gauge kinetic function 
becomes $f(S)=k_\alpha\,S$ where $\alpha$ 
labels $U(1)_Y$, $SU(2)_L$, $SU(3)$.
Note that  we made no assumption about the hypercharge  normalization,
which is arbitrary\footnote{Anomaly cancellation fixes the quantisation of the
 hypercharge, but not its normalization (such as the $3/5$ factor),
 fixed for example  by the presence of a GUT group 
 $SU(5)$, etc, subsequently broken to SM group.}.
A GUT-like normalization for it would actually
demand the ratios in (\ref{sm2}) to be
$b_{A,1}/(5/3)= b_{A,2}=b_{A,3}$. In 
\cite{Chamseddine:1995gb} this was attempted
for unification purposes and link with the heterotic string theory, etc. 
However, since it is difficult to derive a gauged 
$U(1)_R$ from the  heterotic string, one may find this too restrictive 
in some models. Another reason  not  to impose this demand\footnote{See 
 \cite{ky} for non-standard Kac-Moody levels,
and the models  with branes at singularities  \cite{ls1,ls2,ls3}.}
is because in such case anomaly cancellation conditions
via Green Schwarz are not satisfied \cite{Chamseddine:1995gb}.

Eqs.(\ref{sm2}) 
have implications for the  tree level gauge couplings of the SM group
which are  now fixed by  $1/g^2_\alpha\equiv k_\alpha \Re[s]$. 
Let us then estimate the values of couplings if $Q=0$
without GUT-like unification conditions.
Then, with  $\xi=-b\,\cgs$, $b<0$ (see
Sections~\ref{section2.2} and \ref{section2.3})
\footnote{$\xi$ depends on the  model, in
Section~\ref{sec:vacuum} $\xi\!=\!-b\,\cgs$, (see  $V_D$), similar in 
Section~\ref{section2.2}, also \ref{section2.3} if $\beta=0$.}   
eq.(\ref{sm2}) gives positive gauge couplings and
$11/k_1=5/k_2=3/k_3=8\pi^2/(-b)$. 
However, the values of the couplings are not realistic.
With  $4\pi/g_\alpha^2=4\pi\,k_\alpha \Re[s]=
-0.09\times k_\alpha/b$, one finds that
 $4\pi/g_1^2\approx 0.16$,  $4\pi/g_2^2\approx 0.07$, 
 $4\pi/g_3^2\approx 0.04$, so all couplings  are non-perturbative
(for comparison $1/\alpha_{GUT}\approx 25$). 
Therefore\footnote{There are examples where one can lift the GUT-like 
relation {\it and} perturbativity  and still make predictions,
via infrared fixed-point(s) dynamics for  {\it ratios} of these couplings,
which replace the GUT-like constraints  \cite{dg1}.},
 one must consider the case with  non-zero $Q$'s
for the fermions, which allows a perturbative solution if 
$C_\alpha\equiv \Tr[Q\,\tau^a\tau^b]_\alpha\geq \pi\vert c\vert/\vert \Re[s]\vert$, 
where $G_\alpha$ are the SM sub-groups. 
In conclusion, even {\it without} GUT-like unification constraints, these
mixed $U(1)_R$-SM group anomaly cancellation conditions bring strong
constraints. This is so already before considering the $U(1)_R$ cubic 
and $U(1)_R$ mixed-gravitational anomalies constraints.
 This ends the discussion about anomalies in 
which SM subgroups are involved\footnote{The anomaly
$U(1)_R^2-U(1)_Y$  vanishes,  $\Tr Y=0$ on MSSM matter fermions 
(SM-gravitational anomaly).}.

Next, let us consider   the cubic  anomaly of $U(1)_R$. We consider the MSSM spectrum
but also include an additional hidden sector state (fermion) $\psi_z$ of 
$U(1)_R$ charge\footnote{Its superpartner is a  scalar field $z$, of charge $q_z$, see 
also the discussion in Section~\ref{section2.3}.} $R_z$,  which thus  does not affect the discussion 
so far on anomalies involving the  SM group.
Then
\smallskip
\bea
\Tr \, R_\psi^3 & =& 
 3\, ( 2 l^3+e^3+ 6 q^3+
3 u^3+3 d^3)
+2 \,(\tilde h_1^3+ \tilde h_2^3)
+ (8+3+1) (-\xi/2)^3
\nonumber\\
&+&
3\,(-\xi/2)^3+(\xi/2)^3+(-\xi/2)^3
+R_z^3
\nonumber\\
&=&
49\,(Q+\xi/2)^3+15 \,(-\xi/2)^3 +R_z^3
\eea

\medskip\noindent
The first line is due to MSSM matter fermions\footnote{Their 
anomaly contributions are identified by their name in 
a standard notation.} 
and in the second step the  $U(1)_R$   charges were replaced by $Q+\xi/2$. 
The sum $8+3+1$ is due to gauginos of $SU(3)\times SU(2)_L\times U(1)_Y$, 
in this order.
In the second line the first term is due to the
gravitino (3 times the contribution of a gaugino),  the second term 
is due to dilatino, the third term to $U(1)_R$ gaugino and the last
one to the  hidden sector  ($\psi_z$).
We thus find the following condition from eq.(\ref{fieldtheory})
\medskip
\bea
\label{ff}
49 \,(Q+\xi/2)^3+15 \, (-\xi/2)^3+R_z^3
=
\cgs\,(b_C+\xi\,b_{CK}+\xi^2\,b_K)
\eea

\medskip\noindent
For the  mixed $U(1)_R$-gravitational anomaly
we have under similar assumptions for the charges
\bea
\Tr R_\psi&=& 3\,(2 l+ e^c+ 3 u^c + 3 d^c + 6 q) + 2(\tilde h_1 +\tilde h_2)
+
(-\xi/2)\,(13-21)+
(\xi/2) + R_z
\nonumber\\
&=&
49\, (Q+\xi/2)+9 (\xi/2)+R_z
\eea

\medskip\noindent
In the first line
$13=(8+3+1)+1$ from all SM gauginos and $U(1)_R$-gaugino while  $+(\xi/2)$ is due to dilatino.
Then
\bea
49\,Q+29 \,\xi+R_z &=& - 8 \,\cgs \,b_R\label{hh}
\eea
From 
(\ref{ff}), (\ref{hh}), with  $b_C=12\pi^2$,
for canonical gauge kinetic term of $U(1)_R$
one must satisfy
\medskip
\bea\label{ac}
R_z^3 & =& 12 \,\pi^2\,\cgs [1+\xi\,(b_{CK}+\xi\,b_K)/(12\pi^2)]
-49\,(Q+\xi/2)^3+15 (\xi/2)^3
\nonumber\\
b_R & =& -(29\,\xi+R_z+49\,Q)/(8\,\cgs);
\eea

\medskip\noindent
For fixed $Q$, $\xi$, $b_K$ and $b_{CK}$ one 
should adjust $R_z$ and $b_R$ according to eq.(\ref{ac}). 
Thus, canceling the cubic anomaly is possible 
by adding $\psi_z$ of  freely  adjustable  $R_\psi$, even if $b_K=b_{CK}=0$. 
This condition can impact on the existence or stability of the ground
state of the model. Indeed, the scalar $z$ superpartner of $\psi_z$ 
participates in the minimisation conditions of the scalar
potential that fixes the ground state; these
 may impose restrictions  on  $q_z=R_\psi-\xi/2$  (such as its sign) 
inconsistent with the above result for $R_\psi$ \footnote{Recall 
the constraints $q_z<0$ and $\cgs>0$ in Section~\ref{section2.3}, see also  case (b) later on.}. 
These can alter the previous predictions for
$m_{3/2}\sim$TeV. 
One can avoid this case by tuning  $b_K$ or $b_{CK}$ and $b_R$
to respect (\ref{ac}), see eq.(\ref{zz}).

\bigskip\noindent
{\bf  Case (a): no extra state $(z,\psi_z)$}
What happens if no extra hidden state $\psi_z$ is present?
  From eq.(\ref{ac}) with $Q=b_{CK}=b_K=0$
we have $12\,\pi^2\,\cgs=34\,(-b\,\cgs /2)^3$ giving\footnote{
We set $Q\!=\!0$ since with $U(1)_R$ charged MSSM
scalar fields, this value of  $m_{3/2}$ is not valid anymore.}
\bea\label{ou}
\vert m_{3/2}\vert
=  \kappa^{-1}\,
\vert  a \,b/\alpha\,\vert\,e^{\alpha}
\sim
 \kappa^{-1} \vert b\vert^{3/2}  \cgs\,e^\alpha/\vert\alpha\vert
\sim 
(48\pi^2/17)^{\frac{1}{2}} \,\,\kappa^{-1}\,\,e^{\alpha/2}/\vert\alpha\vert.
\eea
Therefore,  the
gravitino mass  becomes of the order of Planck scale and 
is not ``tunable''  anymore to a TeV value. 
As a result, the soft terms masses  would
also become of the order of the Planck scale.
The reason for this result is that the GS mechanism (related to
$\sim \cgs$)  and anomaly cancellation
in the presence of FI terms (related to $\xi\sim b\,\cgs$), 
when  put together are too restrictive given
the  minimal field content in the hidden sector.

\bigskip
\noindent
{\bf Case (b): including the state $(z,\psi)$}:\,\,\,
This is similar to the  model of Section~\ref{section2.3}
``upgraded'' in the visible sector by the MSSM superfields.
These do not alter the discussion there regarding the ground state, etc, 
since the  MSSM scalars are $U(1)_R$ neutral if we set their $Q=0$.
Condition (\ref{ac}) for the cubic anomaly can be re-written as follows
(with $b_{CK}=b_K=0$):
\bea\label{fcb}
(q_z-b\,\cgs/2)^3=12\pi^2\cgs +34 \,(b\,\cgs/2)^3
\eea

\medskip\noindent
using that $R_z=q_z+\xi/2$ with $q_z$  the charge of
the scalar superpartner of $\psi_z$ and $\xi=-b\,\cgs$.

We found in Section~\ref{section2.3} 
that if  $q_z\!<\!0$, $\cgs\!>\!0$  one preserves
the usual ground state and that $\cgs$ is numerically 
very small $ \cgs\! \sim\! 10^{-13}$, $b\!\sim\! \cO(10^{-3})$ 
for a TeV gravitino mass (see  Section~\ref{sec1}).  
This means 
that we can  ignore the last term in the rhs of eq.(\ref{fcb}).
Then the above equation has no solution $q_z\!<\!0$. Therefore,
while one can always add  hidden sector states to cancel  anomalies, 
the result is that the ground state is modified so the 
prediction $m_{3/2}\sim$ TeV is not valid anymore\footnote{
In  Section~\ref{section2.3}, the  gravitino mass was
$\vert m_{3/2}\vert=
\vert \,\kappa^2 \,e^{\kappa^2\,\cK/2}\,W\,\vert
=  \kappa^{-1}\,
\vert  a \,b/\alpha\,\vert\,e^{\alpha}
 \sim
 \kappa^{-1} (-b)^{3/2}  \cgs\,e^\alpha/\vert\alpha\vert$.}
and the phenomenological motivation is lost.

There is in principle one option left: 
 use either  $b_K$ and/or $b_{CK}$ to enforce $q_z<0$ and maintain 
the TeV-value of $m_{3/2}$. This option can also be used for the 
minimal model in Section~\ref{sec:vacuum} to relax its cubic anomaly  constraint in eq.(\ref{hu}).
It is also possible that when adding more fields in the hidden sector of different charges,
the anomalies cancel without changing  the ground state, with
$m_{3/2}\sim~$TeV. But then finding the ground state and its properties
become  difficult tasks.

To conclude, anomaly cancellation in the presence of FI terms and $U(1)_R$ gauge
symmetry, even in the presence of a Green-Schwarz mechanism and after
 relaxing the GUT-like constraints for tree level gauge couplings, 
 is a very restrictive  constraint for model building
(MSSM-like models).  
In an anomaly-free model, a tunable, TeV-scale gravitino mass
may remain possible provided that the $U(1)_R$ charges of additional hidden sector fermions
(constrained by anomalies) do not conflict with the related values of the
 $U(1)_R$ charges of their scalar superpartners, constrained by existence of
 a stable ground state. This issue may be bypassed by tuning instead the 
coefficients of the Kahler connection anomalies ($b_K$ , $b_{CK}$)\footnote{There 
is a possible correction to our analysis
that may be  worth  investigating.
The presence of a non-trivial dilaton  Kahler potential 
$\cK\sim - 2  \ln (s+\overline s)$ lead to a non-flat field  space metric 
$\cK_s^s= \partial^2 \cK/\partial s\partial s^\dagger\sim 2/(s+\overline s)^2$.
The  anomaly cancellation conditions eqs.(\ref{an}), which 
lead to the familiar field theory  
condition on the R-charges (eq.(\ref{ee})) of cubic 
anomaly cancellation, does not take into account the effect of this non-flat metric.
This effect impacts on anomaly cancellation
via a tensor $\Sigma_{\mu\nu\,s}^s=1/(s+\overline s)^2\times (D_\mu s \, D_\nu \overline s-
D_\nu s\, D_\mu \overline s)$, which ``mixes'' the space-time indices with the
field indices. This tensor is just the target space curvature tensor 
``pulled back'' to space-time and is present in the covariant derivatives of the fermions.
In principle, the formalism in \cite{F1,F2}  could be applied to see  if the non-flat 
metric impacts on the cubic and the other anomalies cancellation.}.

\section{Conclusions}
\label{conclusions}

In this work we analyzed, at the classical level,
some models with  a shift symmetry of the dilaton
that is gauged into a  $U(1)_R$ symmetry in 4D $N=1$ supergravity.
We then studied the impact of quantum constraints such as
anomaly cancellation on these models.

At the classical level, a gauged $U(1)_R$ symmetry dictates
the structure of the superpotential $W\sim e^{b\,s}$ where $s$ is the dilaton
which transforms non-linearly under $U(1)_R$.
With a minimal supergravity 
spectrum containing the dilaton (sgoldstino), gravitino, dilatino,
massive $U(1)_R$ gauge boson and its R-gaugino superpartner,
such a toy model can have spontaneous breaking of supersymmetry with
a (small, positive) tunable cosmological constant, TeV-gravitino mass
and gravity-mediation scale.

We showed that these nice properties can be maintained 
in the presence of additional states in the visible and hidden sectors,
under  some  minimal assumptions.
The visible sector can be that  of the MSSM if its chiral superfields
are  considered R-neutral. This means that 
fermions have $U(1)_R$ charges of order $\xi\sim b\,\cgs$, 
where $\xi$ is the Fayet-Iliopoulos constant and  $\cgs$ 
is the shift of the axion, $\Im[s]$.
Additional R-charged field(s) in the hidden sector can be present
and still maintain $m_{3/2}\sim$ TeV, under some constraints for the 
 R-charge(s).

At the quantum level,  we examined the  anomaly cancellation conditions
in supergravity with a gauge group  of SM$\times U(1)_R$.
The spectrum is that of minimal 4D N=1 supergravity
extended by MSSM-like superfields also charged under $U(1)_R$.  
Cubic and mixed anomalies of an anomalous $U(1)$ 
with the SM gauge group, Kahler connection  and gravity were studied in
 the past in a general approach in gauged  supergravity 
with Green-Schwarz mechanism and FI terms. 
We showed the agreement of  these anomaly cancellation conditions 
(other than that involving Kahler connection)  to the ``naive'' field 
theory approach in global SUSY (using Tr over charges and GS mechanism) 
for the special case of gauged $U(1)_R$, with R-charges determined 
using simple arguments (Appendix~A).
Note that in global SUSY $U(1)_R$ cannot be gauged and the 
$U(1)_R$ charges depend on the FI terms.

We then applied the anomaly conditions to  the MSSM as visible sector, with
superfields of  $U(1)_R$-charges equal to $Q$ while fermions charges 
are shifted by the FI  term, and  relaxed the  GUT-like unification condition.
Even after doing so, the $U(1)_R$-SM mixed-anomalies cancellation 
remains a strong parametric constraint that impacts on perturbativity 
of the gauge couplings ($Q=0$). Even without these constraints,  
the cancellation of 
the $U(1)_R$ cubic anomaly  on its own brings constraints on the  gravitino 
mass (and thus  soft terms masses) which becomes of the order of Planck scale.
The addition to the hidden sector of a $U(1)_R$-charged superfield 
does not immediately solve this problem. This is because the anomaly-induced 
constraint on the R-charge of the scalar component
can modify the ground state of the model, thus loosing its properties like 
$m_{3/2}\sim$ TeV. A possible tuning of  the coefficients ($b_K , b_{CK}$) 
of the Kahler connection  anomalies may  bypass this problem.  
Alternatively  a more complicated hidden sector could  avoid this issue, 
but it makes very difficult an analysis of the existence  
of the ground state, with $m_{3/2}\sim$ TeV.

\bigskip\medskip
 \noindent
{\bf Acknowledgements: }

\medskip\noindent
The work of D.~Ghilencea  was supported by  a grant of the 
National Authority for Scientific Research, CNCS-UEFISCDI,
Romanian  Research Council   project number PN-II-ID-PCE-2011-3-0607.
D.G. thanks Hyun Min Lee (Chung-Ang University, Seoul)
for  interesting discussions on this topic.

\newpage

\section*{Appendix A: Gauged $U(1)_R$ and fields charges}
\label{appendixA}
\def\theequation{A-\arabic{equation}}
 \def\thesubsection{A}
 \setcounter{equation}{0}

The action considered is 
 \medskip
\bea\label{yyy}
\cS\!\! &=&\!\!\!\!\!\int
d^4x\,
\Big\{
 d^4\theta\,\bE\,
\Big[  
(-3/\kappa^2)\, S_0^\dagger \,e^{2 \,(\xi/3)\,V_R} S_0\,e^{-\kappa^2\, K_0/3}\Big]
+
\Big[\int d^2\theta \,\cE\,
 S_0^3\,W(\Phi^i)
+
{\rm h.c.}\Big]\Big\}
\nonumber\\[4pt]
\kappa^2\,K & \equiv& \kappa^2\, K_0- 2\,\xi\,V_R
\eea

 \medskip\noindent
$\Phi^i$ are matter superfields charged or not under these groups.
  $S_0$ is the conformal compensator superfield,
$\bE$ is the superspace measure, $\cE$ is the chiral superspace measure.
 $V_R$ is a $U(1)_R$ vector superfield,
$\xi$ is the constant Fayet-Iliopoulos term. 
In the flat limit $\cS\supset \int d^4\theta \,(K_0-2\xi\,V_R)$. If 
$K_0\supset \Phi^\dagger \exp(2\,q\,V_R)
\Phi$ then $\cS\supset q\vert \phi\vert^2-\xi \,D+D^2/2$ giving $D^2\sim (q\vert\phi\vert^2-\xi)^2$.
The action is invariant under a super-Weyl symmetry (eq.(2.9) in \cite{KL})
 \medskip
\bea\label{pp}
\lambda\ra e^{-3\tau}\lambda
\quad
&&
\bE\ra e^{2\tau +2\overline\tau}\,\bE 
\quad
\qquad \cE\ra e^{6\,\tau}\,\cE
\qquad
\cW_\alpha\ra e^{-3\tau}\,\cW_\alpha,
\nonumber\\
V^{(a)}\ra V^{(a)}
\quad
&& 
S_0\ra e^{-2\,\tau}\,S_0
\qquad 
\quad W\ra W
\qquad\,\,\,
\overline\cW^{\dot\alpha} \ra e^{-3\overline \tau}\,\overline \cW^{\dot\alpha}
\eea

\medskip\noindent
with complex parameter $\tau$. Note that the superpotential does not transform, while in
our case it does (so super-Weyl transformation is a particular case
of an $R$-symmetry). We thus need an extra $U(1)_R$; under a $U(1)_R$
gauge transformation 
\footnote{This  transformation on $V_R$
corresponds to a gauge transformation
$A_\mu\ra A_\mu+\partial_\mu \rho$ where $\rho=\Re\Lambda_{\theta=\otheta=0}$.
Under this gauge transformation, the scalar  fields transform as:
 $\phi\ra \exp(-i\, q\, \rho)\phi$, 
with $D_\mu\phi\equiv (\partial_\mu \phi+ i\, g\, q\,A_\mu)\phi$ where $q$ is the charge of 
the field $\phi$. This is consistent
with conventions in the text: $D_\mu \phi^j\equiv \partial_\mu \phi^j-X^j(\phi)\,A_\mu$,
and  $\delta \phi^j=\rho \, X^j(\phi)$ and the Killing vectors
$X^j(\phi)=- i \,q \,\phi^j$ for a linearly transformed $\phi$. 
These  conventions are similar to those in \cite{V1} but charges $q$ and $\xi$
of  opposite signs.}
\bea
V_R & \ra &  V_R+\frac{i}{2}\,(\Lambda-\Lambda^\dagger),\qquad \overline D\Lambda=0
\eea
$K$ must then transform ($K_0$ invariant)
\bea
\kappa^2\,K  & \ra & \kappa^2\,K -  i\,\xi\,(\Lambda-\Lambda^\dagger)                
\eea
The action is invariant under $U(1)_R$ if
\bea
S_0\ra S_0^\prime = e^{ - i \,\xi/3\,\Lambda}\,S_0,
\qquad
W\ra W^\prime=e^{+ i\,\xi \Lambda}\,W\qquad \Ra \,\,\,\, R_W=-\xi.
\eea
and  $G = \kappa^2\,K+\ln \vert W\,\kappa^{3}\vert^2$ is invariant, too\footnote{
With $W\!=\!a\,e^{b\,S}$, ($b<0$), under transformation $S\ra S-i\,\cgs\,\Lambda$ gives
$\xi=-\, b\,\cgs$ and $D^2\!\sim \!(q\vert\phi\vert^2\!+b\,\cgs)^2$.}.
The choice of super-Weyl gauge $S_0=s_0+\theta^2\,F$ can be maintained if we combine the previous 
super-Weyl and the $U(1)_R$ transformations such as the conformal compensator remains invariant
(neutral). This is possible provided that \cite{Lee:2008zs}
\bea
-2\tau - i\,(\xi/3)\,\Lambda=0,\quad \ra \quad \tau=-\frac{i\,\xi \Lambda}{6}.
\eea

\medskip\noindent
thus 
$R_\lambda=R_{\cW}=-\xi/2$
 according to (\ref{pp}).
We thus work in this gauge which keeps manifest SUSY and holomorphicity
 and the compensator is neutral under this $U(1)_R$. A consequence
is that, according to the transformation of $\bE$,  the 
gravitino will carry a charge under this new $U(1)_R$.
Further, we also have
\bea\label{R}
R_{V}=0,\,\,\,\,
R_\cW=-\frac{\xi}{2}=R_{\bar D^2 \,D}
\eea

\medskip\noindent
With $V\supset \otheta\otheta\theta\lambda$ then $R_\theta=R_\lambda=-\xi/2$.
A superfield of R-charge $Q$ transforms as $\Phi\!\ra\! e^{- i\,Q \Lambda}\,\Phi$.

The component form of the action contains, 
for a superfield $\Phi=(\phi,\psi_\phi)$ \cite{WB}
\medskip
\bea
L\supset X^\phi\,\overline\psi_\phi\,\overline\lambda+h.c.
\eea

\medskip\noindent
$X^\phi$ is the Killing vector of scalar $\phi$
and $\lambda$ the R-gaugino.
For a  superfield with $\Phi\ra e^{- i\,Q\,\Lambda}\,\Phi$  
\medskip
\bea
X^\phi\propto i\,\phi\qquad \Ra\qquad 
R_{\psi_\phi}=R_{\phi}-R_\lambda=
Q+\frac{\xi}{2}
\eea

\medskip\noindent
which was used in the text. 
Consider now that $K_0$ contains  a dilaton ($S$) dependent term
\medskip
\bea
K_0\supset -\ln(S+S^\dagger-\delta\, V_R)
\eea

\medskip\noindent
which is invariant under our gauged $U(1)_R$ provided
\medskip
\bea
S\ra S+i\,\frac{\delta}{2}\,\Lambda,\quad \delta\,\,{\rm real};
\qquad
s\ra s+i \,\frac{\delta}{2} \,\rho,\qquad
\rho\equiv (\Re\Lambda)\vert_{\theta=\otheta=0}
\eea

\medskip\noindent
For the dilaton $S=(s,\psi_s)$
\medskip
\bea
X^s=+i\delta, \quad\Ra\quad 
 R_{\psi_s}= -R_\lambda=\frac{\xi}{2}\quad \Ra\quad R_{\phi^s}=R_{\psi_s}+R_\theta=0.
\eea

\medskip\noindent
Regarding the gravitino $\psi_{(3/2)}$, from any of the terms of the supergravity Lagrangian 
\medskip
\bea
L\supset -\frac{1}{2}\,e\,D_a\,\psi_{(3/2)}\,\sigma\,\overline\lambda^{a}+
e^{K/2}\,W\,\opsi_{(3/2)}\sigma \opsi_{(3/2)}+h.c.
\eea

 \medskip\noindent
one obtains the value of $R_{\psi_{3/2}}$ used in the text:
\bea
R_{\psi_{(3/2)}}=R_\lambda=\frac{1}{2}\,R_W=\frac{-\xi}{2}.
\eea

\section*{Appendix B: 
Cancellation of anomalies with a  $U(1)_R\times {\rm SM}$  group}
\label{appendixB}
\def\theequation{B-\arabic{equation}}
 \def\thesubsection{B}
 \setcounter{equation}{0}

Consider the Lagrangian in the global SUSY limit, with S the dilaton:
\medskip
\bea
\cL\!=\!
-\!\int\!\! d^4\theta 
\ln(S+\overline S -\delta\,V_R)+
\Big\{\frac{1}{16\pi^2\,\kappa}
\int\!\! d^2\theta
\Big[ k_R \,S\,\cW_R^\alpha \cW_{R,\alpha}
\!+\!
k_a\,S\,\Tr\,\cW^\alpha_a\,\cW_{a,\alpha}\Big]\!
\!+\!{\rm  h.c.}\Big\}\,
\eea

\medskip\noindent
where $\kappa$ cancels the $\Tr$ factor in non-Abelian case. $k_R$ and $k_a$ are
Kac-Moody levels of $U(1)_R$ and subgroups $G_a: U(1)_Y$, $SU2)_L$, $SU(3)$
  of the SM group. For example for $U(1)_R$ part
\medskip
\bea
\cL\supset \frac{1}{16\,g^2}\!\!
\int d^2\theta\,k_R\,S \,\cW_R^2
\supset
k_R \,\Big\{\frac{-1}{4}\,\Re[s]\, C_{\mu\nu}\,C^{\mu\nu}
+\frac{1}{2}\,\Re[s]\,D_a^2+
\frac{1}{4}\,\Im[s]\,C_{\mu\nu}\,\tilde C^{\mu\nu}\Big\}
\eea

\medskip\noindent
where $\tilde C^{\mu\nu}=1/2\,\epsilon^{\mu\nu\rho\sigma} C_{\rho\sigma}$.
Consider the  shift 
\bea
S\ra S+i\,\frac{\delta}{2}\,\Lambda(x)
\eea
and define
\bea
C_\alpha=\Tr_{G_\alpha}\,[T(r)^2\,R_\psi]
\eea
$C_\alpha$ denotes the mixed anomaly $U(1)_R$  with $G_\alpha=U(1)_Y, SU(2)_L$, or  $SU(3)$ and 
$R_\psi$ is the $U(1)_R$ charge with fermions transforming as $\psi\ra \exp(-i\,R_\psi\,\rho)\,\psi$
with $\rho=\Re\Lambda\vert_{\theta=\overline\theta=0}$. $R_\psi$ depend on the Fayet-Iliopoulos
 constant ($\xi$) as discussed in previous Appendix.

The anomalous $U(1)_R$ generates
$\Delta\cL\propto \int d^2\theta [ \,i\,C_\alpha\,\Lambda\,\cW_a\,\cW_{a}]$
for each subgroup $G_\alpha$.
The shift of the dilaton is the same for all $G_\alpha$ (ignoring $k_\alpha$) then
the ratio $C_\alpha/k_\alpha$ must be identical  for all $G_\alpha$ for anomalies to cancel. 
Also taking the $U(1)^3_R$ and $U(1)_R$-gravitational
anomaly, then one has the result
\bea\label{an1}
- 4\,\pi^2\delta=\frac{2\,C_\alpha}{k_\alpha}
=\frac{(2/3)\, \Tr \,R_\psi^3}{k_R}=\frac{1}{12}\,\Tr \,R_\psi,
\eea

\medskip\noindent
In the paper we kept $k_R=1$. 
The gauge couplings constants are  then $k_\alpha\,\langle \Re[s]\rangle\equiv 1/g_\alpha^2$.

Regarding the mixed gravitational anomaly (last term above) this is seen from the action 
\medskip
\bea
L\supset  \frac{-1}{4}\, \Im[s]\,\cR\,\tilde \cR,
\qquad
\Delta L_{axion}=
-\frac{\rho}{8}\,\delta\,\cR\tilde \cR,
\qquad
\Delta L_{anomaly}=
\frac{\rho}{384\,\pi^2}\, \Tr [R_\psi]\,\cR\tilde \cR\,
\eea

\medskip\noindent
and use $\Delta L_{axion}+\Delta L_{anomaly}=0$.
However, unlike in heterotic string, in supergravity,  
  $\cR\tilde\cR$ term can have a different coefficient from $\Im[s]$
since it is not part of the leading order action.

In the text we used the notation 
$W=a\,e^{b\,S}$, with $b<0$ and  $S\ra S-i\,\cgs\,\Lambda$ giving
$\delta/2=-\cgs$. The anomaly condition (\ref{an1}) becomes
\bea\label{an2}
4\pi^2\,\cgs=\frac{C_\alpha}{k_\alpha}=\frac{(1/3)\,\Tr R_\psi^3}{k_R}=\frac{1}{24}\,\Tr[R_\psi].
\eea
This result was compared against the more general 
supergravity results in Section~\ref{anomalies}.


\end{document}